\documentclass[sigconf]{acmart}
\AtBeginDocument{%
  }

\copyrightyear{2025}
\acmYear{2025}
\setcopyright{cc}
\setcctype{by-nd}
\acmConference[CHI '25]{CHI Conference on Human Factors in Computing Systems}{April 26-May 1, 2025}{Yokohama, Japan}
\acmBooktitle{CHI Conference on Human Factors in Computing Systems (CHI '25), April 26-May 1, 2025, Yokohama, Japan}\acmDOI{10.1145/3706598.3714127}
\acmISBN{979-8-4007-1394-1/25/04}

\usepackage{amsmath}
\usepackage{graphicx}
\usepackage{booktabs}
\usepackage{subcaption}
\usepackage[normalem]{ulem}
\useunder{\uline}{\ul}{}
\begin{document}

\title[Breaking the Familiarity Bias]{Breaking the Familiarity Bias: Employing Virtual Reality Environments to Enhance Team Formation and Inclusion}

\author{Mariana Fernandez-Espinosa}
\orcid{0009-0004-1116-2002}
\affiliation{%
  \institution{University of Notre Dame}
  \city{Notre Dame}
  \state{Indiana}
  \country{USA}
}
\author{Kara Clouse}
\orcid{0009-0009-6466-7043 }
\authornote{Authors worked on this project as part of Notre Dame's Interdisciplinary Traineeship for Socially Responsible and Engaged Data Scientists (iTREDS).}
\affiliation{%
  \institution{University of Notre Dame}
  \city{Notre Dame}
  \state{Indiana}
  \country{USA}
}
\author{Dylan Sellars}
\orcid{}
\authornotemark[1]
\affiliation{%
  \institution{University of Notre Dame}
  \city{Notre Dame}
  \state{Indiana}
  \country{USA}
}
\author{Danny Tong}
\orcid{0009-0002-2449-0991}
\authornotemark[1]
\affiliation{%
  \institution{University of Notre Dame}
  \city{Notre Dame}
  \state{Indiana}
  \country{USA}
}
\author{Michael Bsales}
\orcid{}
\authornotemark[1]
\affiliation{%
  \institution{University of Notre Dame}
  \city{Notre Dame}
  \state{Indiana}
  \country{USA}
}
\author{Sophonie Alcindor}
\orcid{0009-0005-8035-2587}
\authornotemark[1]
\affiliation{%
  \institution{University of Notre Dame}
  \city{Notre Dame}
  \state{Indiana}
  \country{USA}
}
\author{Tim Hubbard}
\orcid{0000-0002-6029-7446}
\affiliation{%
  \institution{University of Notre Dame}
  \city{Notre Dame}
  \state{Indiana}
  \country{USA}
}
\author{Michael Villano}
\orcid{0000-0002-5212-326X}
\affiliation{%
  \institution{University of Notre Dame}
  \city{Notre Dame}
  \state{Indiana}
  \country{USA}
}
\author{Diego Gomez-Zara}
\orcid{0000-0002-4609-6293}
\email{dgomezara@nd.edu}
\affiliation{%
  \institution{University of Notre Dame}
  \city{Notre Dame}
  \state{Indiana}
  \country{USA}
}
\additionalaffiliation{%
  \institution{Pontificia Universidad Cat\'olica de Chile}
  \department{Facultad de Comunicaciones}
  \city{Santiago}
  \country{Chile}
}
\renewcommand{\shortauthors}{Fernandez-Espinosa et al.}

\begin{abstract}
Team closeness provides the foundations of trust and communication, contributing to teams' success and viability. However, newcomers often struggle to be included in a team since incumbents tend to interact more with other existing members. Previous research suggests that online communication technologies can help team inclusion by mitigating members' perceived differences. In this study, we test how virtual reality (VR) can promote team closeness when forming teams. We conducted a between-subject experiment with teams working in-person and VR, where two members interacted first, and then a third member was added later to conduct a hidden-profile task. Participants evaluated how close they felt with their teammates after the task was completed. Our results show that VR newcomers felt closer to the incumbents than in-person newcomers. However, incumbents' closeness to newcomers did not vary across conditions. We discuss the implications of these findings and offer suggestions for how VR can promote inclusion.
\end{abstract}




\maketitle

\section{Introduction}
\label{introduction}
Bringing new members to a team is one of the most effective ways to foster originality, innovation, learning, and performance \cite{zeng2021fresh,LEWIS2007159}. Newcomers can bring new perspectives and resources to a group, including knowledge, skills, and social connections \cite{yuan2020making}. Moreover, organizations and groups are not static. Employees will leave due to turnover, promotions, or transfers to another unit. Thus, teams must often recruit and include new members to continue their work and facilitate new ideas \cite{guimera2005}. 

However, incorporating new members into an existing group can be a challenging task \cite{kraut2010dealing}. Newcomers need time to understand the group's dynamics, feel included, and commit to the team. Furthermore, team familiarity (i.e., team members' prior experience working with one another \cite{pasarakonda2023team, muskat2022team}) exacerbates the attachment to previous collaborators and similar individuals in a group, making newcomers' inclusion more difficult \cite{Arcsin2021}. As a result, existing team members (i.e., incumbents) tend to rely on their established relationships, trusting and interacting primarily among themselves. 
 
Previous research in Human-Computer Interaction (HCI) has examined how online communication systems can enhance group dynamics by modifying team members' appearances and facilitating their interactions \cite{harris2019joining,10.1145/2998181.2998300,Whiting2020}. While much work has focused on features that promote inclusion among team members, little has been explored on how these technologies affect newcomers and incumbents differently. Additionally, although previous studies have investigated how online communication systems influence perceived closeness among team members, this has been mostly explored at the interpersonal level, with limited consideration of how these perceptions function within an entire team \cite{Hall2022}. This gap raises several questions about how online communication systems can differently affect the inclusion of new members. Would newcomers be perceived differently depending on the online communication system in which they meet their new team? Would incumbents be more welcoming if the differences between them and the newcomers were less perceptible? 

In this study, we examined how employing a collaborative Virtual Reality (VR) application affects team members' closeness to each other. We investigated VR because its high-level immersion and customizable avatars can potentially reduce the influence of physical appearances, cultural cues, and other factors that often lead to preconceived ideas and negative stereotypes among team members \cite{christofi2017virtual, higgins2023perspective, mal2023impact}. Potentially, collaborative VR could foster a more inclusive atmosphere and allow team members to focus more on the essence of collaboration rather than interpersonal biases \cite{latoschik2017effect}. Moreover, we ask whether using VR can mitigate incumbents' familiarity bias---their tendency to collaborate and interact more with familiar members than with new members. Given our interest in exploring how VR can change the relationships between newcomers and incumbents, our research questions are:
\begin{itemize}
    \item[\textbf{RQ1:}] How does using Virtual Reality affect incumbents' perceived closeness to newcomers?
    \item[\textbf{RQ2:}] How does using Virtual Reality affect newcomers' perceived closeness to incumbents?
    \item[\textbf{RQ3:}] Can Virtual Reality reduce familiarity bias among incumbents toward newcomers?
\end{itemize}

To answer these research questions, we conducted a controlled, between-subject experiment with 29 teams of three members each. Teams were randomly assigned to work in one of two settings: in-person or a VR multi-player application (Meta Horizon Workrooms). In each experimental session, two of the three participants were initially brought together into a room to get to know each other (i.e., the incumbents), while the third participant (i.e., the newcomer) joined later. Teams then completed a hidden-profile task, which required all team members to share information and have a collaborative discussion to succeed. Afterward, participants completed a post-treatment survey to evaluate their work experience and their relationships with their teammates. We employed mixed methods to analyze the survey data, which included behavioral scales and open-ended questions. 

Our findings reveal that using VR significantly impacted team members' closeness only for newcomers. Newcomers in VR reported feeling closer to their incumbents than the newcomers working in-person, and this effect was mediated by the high levels of perceived similarity experienced in VR. However, the incumbents in VR were not affected in the same way, and employing VR did not cause any significant differences in incumbents' familiarity bias toward newcomers. Through thematic analysis \cite{braun2006using}, we found that the VR setting provided participants with psychological safety by reducing social pressures, while the immersive virtual environment enhanced participants' sense of presence and engagement in the collaborative task. 

This paper provides the following three contributions. First, it deepens our understanding of the asymmetric effects of VR on team members' closeness and familiarity perceptions, as newcomers can feel closer to their team, while incumbents may not experience any differences. Second, it offers a controlled between-subject experiment comparing team members' closeness in VR and in-person settings, including a de-identified dataset available at OSF.io \cite{Gomez-Zara_2025} to facilitate reproducibility and further research. Third, it discusses how the quantitative and qualitative results can enhance future VR applications for teams and collaborations, elaborating on design implications that promote team members' closeness and inclusion.

\section{Literature Review}
\label{literature_review}
We situate our work in the context of prior studies of teams---including closeness and familiarity---and VR research.

\subsection{Teams}
Teams consist of two or more individuals collaborating to achieve a shared objective \cite{salas2000teamwork}. As the foundation of organizational work, teams coordinate efforts to tackle complex tasks, demonstrating interdependencies in workflows, goals, and outcomes that reflect their collective responsibility \cite{kozlowski2006enhancing, gomez2020taxonomy, baron2003group}.

\subsubsection{Team Closeness}
Team closeness measures team members' subjective perception of how connected they feel to other members of the team \cite{Gachter2015}. Individuals can feel closer to specific team members based on their interactions and not necessarily on their physical proximity \cite{Wiese2011}. While team cohesion usually refers to how united team members are as a whole \cite{salas2015measuring}, closeness relates to relationships between pairs, which has been studied at the interpersonal level \cite{rosh2012too}. 

Researchers have identified several factors that contribute to building closeness \cite{moreland1982exposure}. Time is a key element, enabling teams to build shared knowledge and establish trust among members \cite{harrison2003time, gillespie2012factors, cattani2013tackling}. Previous relationships also play an important role, as positive prior experiences can reduce uncertainty among interactions between team members \cite{de2017attuning, jones2019essentials, dittmer2020cut}. Lastly, similarity affects closeness since team members who share common attributes such as demographics, values, or experiences are more likely to develop stronger connections and a sense of familiarity \cite{hinds2000choosing, ruef2003structure, winship2011homophily}.

\subsubsection{Team Familiarity}
Team familiarity refers to the amount of experience they have working together \cite{muskat2022team,mukherjee2019prior}. Previous research has shown that team familiarity can be a predictor of improved performance, enhancing creativity, efficiency, and the overall quality of a team's output \cite{dittmer2020cut, huckman2009team, witmer2022systematic}. For instance, Sosa demonstrated through a sociometric study that teams with stronger interpersonal connections and shared knowledge bases were significantly more likely to generate innovative solutions to complex problems \cite{sosa2011creative}. Similarly, Staats \cite{staats2012unpacking} found that increased familiarity was associated with improving team performance. Lastly, Salehi et al. \cite{10.1145/2998181.2998300} found that crowd worker teams performed better when some of their members had previously worked together. 

While its benefits have been well documented, team familiarity also produces adverse effects. It can lead to a culture where incumbents develop close networks and pose significant barriers to new members, such as exclusion \cite{choi2004minority, joardar2007experimental}, intimidation \cite{topa2016newcomers}, favoritism \cite{balthazard2006dysfunctional}, or communication barriers \cite{kraut2010dealing}. Prior work shows newcomers are often perceived as less capable or influential, limiting the team's ability to explore diverse perspectives and solutions. Moreover, high levels of team familiarity can discourage individuals from challenging established viewpoints, ultimately hindering knowledge sharing and the generation of innovative ideas \cite{assudani2011role, xie2020curvilinear}.

\subsubsection{Newcomers}
Including newcomers to a team is important for organizations. Newcomers often bring fresh ideas, innovative approaches, and valuable resources that are crucial for the ongoing vitality of the group \cite{zeng2021fresh}. However, integrating newcomers effectively into an established group poses considerable challenges. They may face strong biases, and their mere presence, even when adhering to social norms, can be perceived as disruptive by existing members \cite{kraut2010dealing, spertus2001scaling}. These issues can make the existing group less desirable for those familiar with the existing dynamics. For example, Joardar et al. \cite{joardar2007experimental} highlighted that the introduction of newcomers often triggers resistance from incumbents who may struggle to accept them as part of the group, potentially leading to tension and disruption of effective team functioning. Furthermore, cultural similarity between the newcomer and the incumbents plays a crucial role in facilitating positive group acceptance. Newcomers facing a significantly different socio-cultural environment are often less familiar with local expectations and norms, which can impede their ability to join smoothly \cite{furnham1982social}. The lack of cultural similarity can exacerbate the groups' challenges, complicating the newcomers' acceptance and diminishing their team cohesion \cite{nesdale2000immigrant}. 

\subsubsection{Online Group Communication}
The new settings of remote, hybrid, and in-person (IP) workplaces are reshaping how individuals perceive themselves and assess their relationships with co-workers. Prior research highlights that online communication systems affect individuals' perceived differences and connections within groups \cite{Hall2022,DESCHENES2024100351}. Since individuals are not physically together, systems' design factors such as nonverbal cues, facial expressions, members' representation, and synchronous conversation will influence how aware participants are of their differences and similarities \cite{giambatista2010diversity}. Consequently, online communication systems can moderate how much rich information members can have of each other, leading users to perceive less of their demographic and physical differences \cite{carte2004capabilities,SUH1999295}. Providing more information about their characteristics and differences has the risk of avoiding teaming up with others who are different or unfamiliar \cite{gomez2020impact}. 

HCI research has shown that online technologies can mitigate biases toward familiar individuals by altering the presentation of user identities \cite{maloney2020anonymity, Whiting2020, Tzlil2018}. Several online platforms offer mechanisms like pseudonyms, avatars, and anonymity to depersonalize interactions \cite{Shemla2016}. These features reduce the emphasis on personal details, enabling more equitable participation in discussions and decision-making processes, which can lead to fairer and more balanced outcomes \cite{christopherson2007positive, joinson2001self, walther1996computer}. In particular, anonymity allows users to create online personas distinct from their offline identities, fostering self-expression and experimentation while promoting inclusivity and reducing biases in collaborative and social settings \cite{bargh2002can, yurchisin2005exploration, kang2013people}. However, it may also lead to undesired negative behaviors, such as group polarization and harassment, due to the reduced sense of accountability \cite{weisband1993overcoming, Ma2016, christopherson2007positive}.

\subsection{Virtual Reality}
VR integrates advanced technologies that create immersive and interactive 3D environments \cite{wohlgenannt2020virtual, 10.1145/3613905.3651085, hubbard2024cross}. By closely emulating the dynamics of IP conversations, VR enables synchronous and embodied interactions, allowing users to engage with the virtual space and each other through natural body movements and vocal communication \cite{laato2024making, abbas2023virtual, mccloy2001science}. In recent years, many VR applications have increasingly supported collaborative interactions among multiple users, enabling them to be in the same virtual space regardless of their physical locations \cite{li2021social,10.1145/3411764.3445426}. Previous research has explored several features that enhance group interactions in VR, including the use of avatars, the sense of social presence, and the spatial configuration that VR can provide \cite{10.1145/3584931.3606992, sykownik2021most, fang2023towards}. 

Avatars provide users with the flexibility to choose how they represent themselves in VR, serving as the primary identity cue that shapes perceptions and interactions \cite{waltemate2018impact}. This representation influences user behavior through stereotypes, a phenomenon known as the ``Proteus effect'' \cite{yee2007proteus}. For instance, research in VR demonstrates that embodying an elderly avatar in immersive environments can reduce stereotypical attitudes toward older individuals \cite{yee2006walk}. Similarly, Groom et al. \cite{groom2009influence} found that using avatars of a different race in VR led to measurable shifts in racial attitudes, suggesting that virtual embodiment enables users to challenge biases through flexible and diverse self-representation \cite{banakou2016virtual}. Avatar-based interactions that mimic non-verbal cues foster these social interactions \cite{freeman2016intimate, clark1991grounding, oh2016let}. Maloney et al. \cite{maloney2020talking} found that non-verbal communication in social VR was perceived as positive and effective, as it offered a less intrusive method for initiating interactions with online strangers. Similar research by Xenakis et. al \cite{xenakis2022nonverbal} highlighted how the reduction of body signals in VR, such as facial expressions and gestures, can decrease reliance on traditional physical cues. 

Prior research has also demonstrated that social presence is a critical factor for effective collaboration in VR \cite{kimmel2023lets, sterna2021psychology, yassien2020design}. VR users can experience an enhanced sense of co-presence, which creates the illusion of sharing a virtual space with their colleagues. The immersive quality of VR helps reduce the sense of detachment often associated with remote work, fostering more engaging interactions \cite{wienrich2018social, smith2018communication} and driving users' intentions to collaborate \cite{mutterlein2018specifics}.

Lastly, the spatial configuration of VR environments impacts how team members recognize and interact with each other. Proxemics, or spatial behavior, refers to the measurable distances between people as they communicate, influencing interpersonal dynamics \cite{hans2015kinesics}. In a virtual setting, virtual proximity replicates this effect, enabling team members to experience a shared presence in a virtual room, even if they are physically distant \cite{williamson2021proxemics}. This shared virtual space fosters a sense of connection and engagement, helping users recreate the interpersonal dynamics, such as proximity and orientation, that are important for effective collaboration \cite{li2021social, williamson2022digital}. 

VR environments influence individuals' perceptions and behaviors during collaboration by abstracting physical features and reducing appearance-based cues that often highlight differences among team members. This shift redirects attention toward collaborative interactions, potentially diminishing social biases and the perceived distinction between incumbents and newcomers. Building on this reasoning, we propose that the mode of interaction affects both how incumbents perceive closeness to newcomers and how newcomers perceive closeness to incumbents. Our first two hypotheses are:

\begin{quote}
    \textbf{H1}: \textit{When onboarding new members, incumbents will perceive greater closeness to newcomers when meeting in VR than when meeting in person.}
\end{quote}
\begin{quote}
    \textbf{H2}: \textit{When onboarding new members, newcomers will perceive greater closeness to incumbents when meeting in VR than when meeting in person.}
\end{quote}

When individuals unfamiliar with one another come together to assemble a team, they initially have limited information to assess each other or establish expectations for their interactions. To alleviate uncertainty, team members often rely on readily available cues, such as physical appearance or demographic traits, to quickly form impressions \cite{allport1954nature,hinds2000choosing}. However, these initial judgments can reinforce social biases, favoring familiar individuals or strengthening preexisting group distinctions. Prior research suggests that VR can reduce such biases through depersonalization and flexible identity representation. By shifting attention away from physical and cultural differences as well as preexisting relationships, VR can foster a more inclusive environment that reduces perceived distinctions between incumbents and newcomers. Therefore, we hypothesize:
\begin{quote}
    \textbf{H3}: \textit{Incumbents' familiarity bias against newcomers will be lower when meeting in VR than when meeting in person.}
\end{quote}

Lastly, one mechanism that could explain how VR fosters closeness is \textit{perceived similarity}---team members who see themselves as more alike tend to experience higher trust and a stronger sense of bonding \cite{zellmer2008and, mannix2005differences}. In VR, avatars and immersive environments can attenuate physical or demographic differences \cite{lopez2019investigating, marini2022can}, potentially making team members feel more similar and accentuating shared goals. This reduction in perceived differences may enhance perceptions of similarity, ultimately fostering closeness. Thus, we hypothesize:
\begin{quote}
    \textbf{H4}: \textit{Perceived similarity will mediate the relationship between VR (versus in-person) and team closeness, such that using VR increases perceived similarity among team members, thereby enhancing closeness.}
\end{quote}

\section{Methodology}
\label{methodology}
We conducted a between-subject experiment with 87 participants to test our hypotheses, assessing the perceived closeness among team members based on the experimental conditions and exploring other factors related to team behavior. 

\subsection{Participants}
The study was approved by the University of Notre Dame's Institutional Review Board (IRB) under protocol number 23-10-8142. We recruited participants through direct outreach by the research team, Notre Dame Psychology Department's SONA system (an online platform used to recruit participants for experiments), and social media channels.

\subsection{Task Description}
All participants were randomly assigned to teams of three people to perform a collaborative task. Since we aimed to test preferences toward incumbents versus newcomers, we randomly asked two participants to meet first and complete an ice-breaker exercise (Appendix \ref{appendix:ice-breaker}). These two participants spent ten minutes getting to know each other. After that, the third member was introduced to the existing two participants, and the research assistant (RA) explained the task to them.

We asked the three participants to complete a task adapted from the ``hidden profile'' paradigm \cite{stasser1985pooling}, which requires all participants to review candidates and select the best one for a managerial position. The information about the candidates is deliberately distributed unequally among team members to create a ``hidden profile,'' where no single individual has all the necessary information to make the optimal choice. This task requires all team members to share their unique pieces of information and integrate them to identify the most suitable candidate. If only a few members shared information or dominated the group decision-making process, the team would not pick the best candidate. As such, this task can reveal biases in group decision-making, such as members dominating the conversation, low participation, and absence of a collective discussion. Hidden profile tasks have been employed in many laboratory experiments to test information sharing between group members when making decisions \cite{Goyal2014,mentis2009,mennecke1997using}. 

In our study, we requested participants to collectively decide on three candidates for the presidency of a university's new satellite campus. Each participant received a copy of the candidates' resumes, each with different information about the candidates' positive and negative attributes. Since each candidate's attributes varied in each resume's version, the participants needed to share and discuss the differing pieces of information provided in their respective copies. One candidate was appropriate to be hired if all participants shared their versions of the candidates' resumes. The task for this study was carefully designed to resemble real-world decision-making scenarios, such as hiring or promotions, and to make them relatable and meaningful to the participants. We instructed participants not to share the provided copies of the resumes with their teammates. The RA also monitored participants' progress by watching them through a video camera in the IP condition and observing the VR meeting room on a computer in the VR condition. 

We chose this task and the candidates' resumes based on three criteria: (a) that the task's content was currently relevant in organizational contexts, (b) that the task did not require extensive training to be completed, and (c) that the participants were likely to engage with the task since they will not know their teammates beforehand. The resumes are available in the Supplementary Materials.

 \begin{figure*}[!htb]
\centering
    \begin{subfigure}[b]{.48\textwidth}
        \includegraphics[trim={0 2.5cm 0 3cm},clip,width=\textwidth]{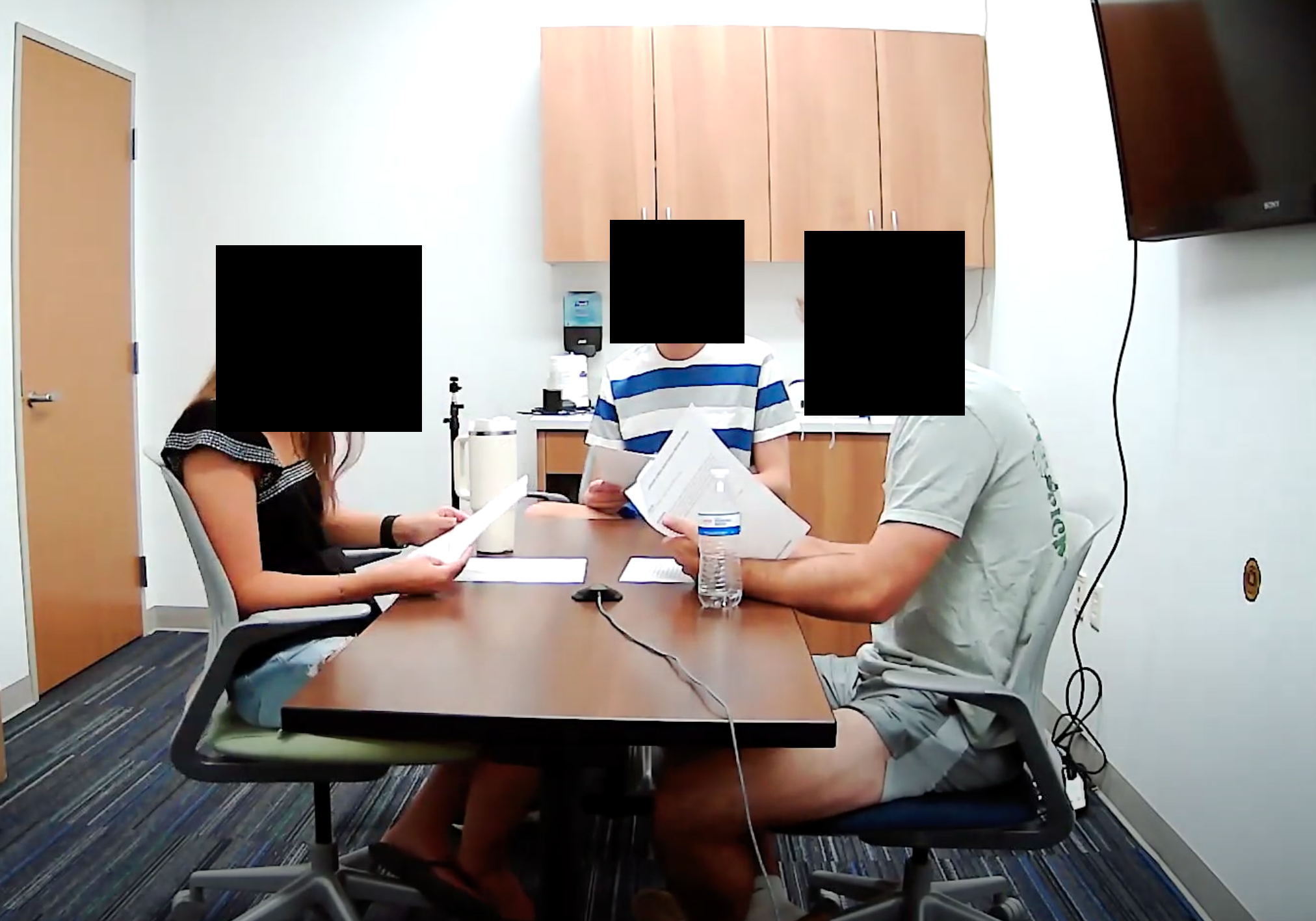}
        \Description{In-Person.}
        \caption{In-Person}
        \label{figure:in-person-session}
    \end{subfigure}\qquad
    \begin{subfigure}[b]{.47\textwidth}
        \includegraphics[width=\textwidth]{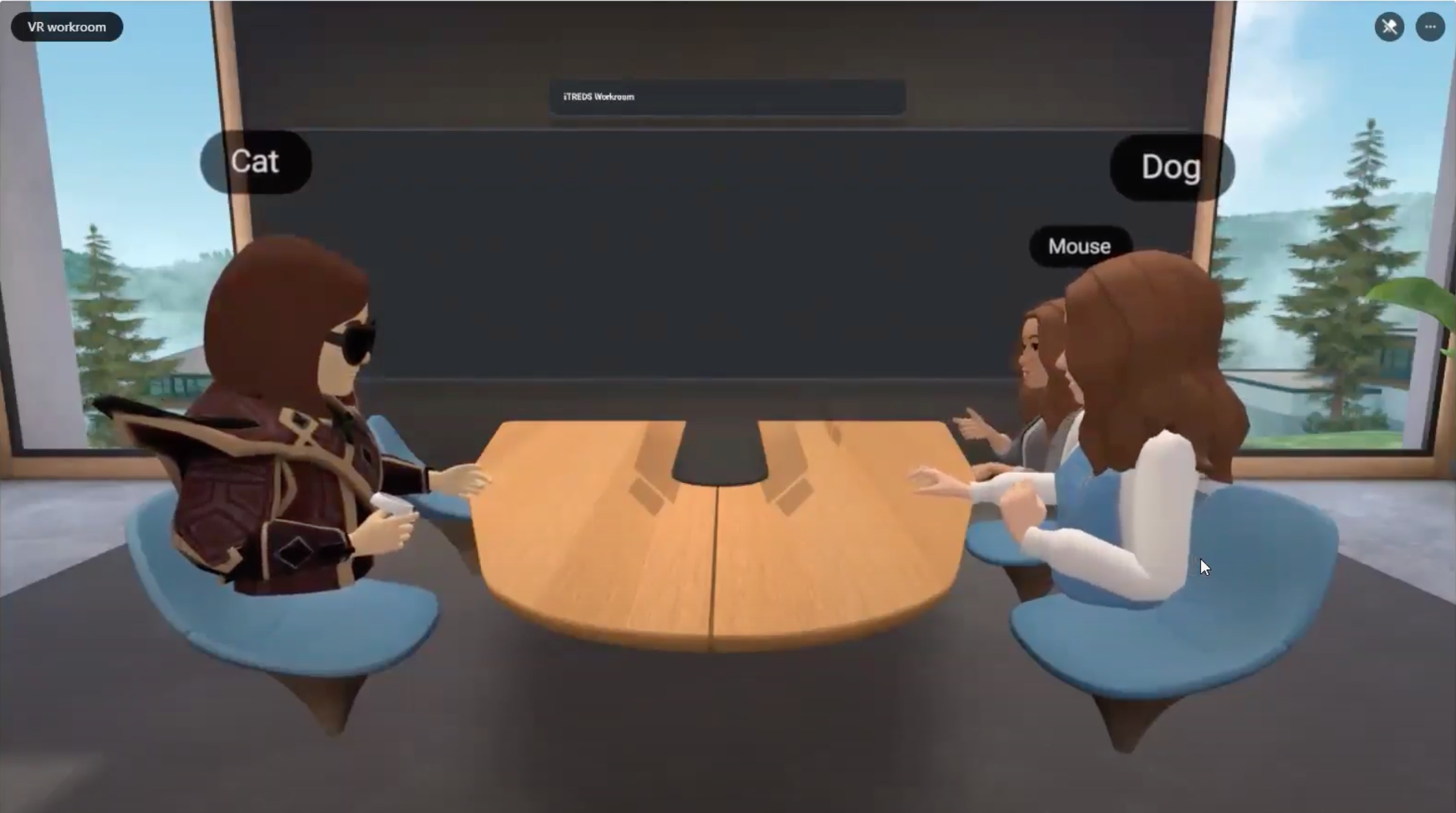}
        \Description{Virtual Reality (VR)}
        \caption{Virtual Reality (VR)}
        \label{figure:vr-session}
    \end{subfigure}
\caption{Experimental Conditions. On the left, participants are working in person. On the right, participants are working using Meta Horizon Workrooms. In both conditions, participants randomly sat in one of the chairs available.}
\Description{Experimental Conditions. On the left, participants are working on the task In-Person. On the right, participants are working on the task using Meta Horizon Workrooms. Participants were randomly assigned to a seat in both conditions.}
\label{fig:experimental-conditions}
\end{figure*} 

\subsection{Conditions}
The teams were randomly assigned to perform this task \textit{In-Person (IP)} or using \textit{Virtual Reality (VR)}. Meeting IP is the most natural communication medium and a ``gold standard'' to compare communication technology \cite{harrison2020framing}. We situated participants in an experimental room with a web camera to record their conversations. Two of the three participants were together in a physical room, while the third participant remained in a separate room. After the two participants completed the ice-breaker activity, the third participant joined them in the same room. Participants randomly sat in one of the three chairs. Each participant had a paper copy of their respective versions of the candidates' resumes, and we asked them not to display the papers to the other participants. All participants sat together at a table, with an omnidirectional microphone to record their discussion and final candidate choice (Figure \ref{figure:in-person-session}).

In the VR condition, all team members were located in different physical rooms with a Meta Oculus Quest 2 headset. The devices had activated hand tracking, enabling users to interact with the application directly using their hands. The RA helped participants wear the headset and learn to control the interface. They had five minutes to create their virtual avatars, and we instructed them to create avatars that looked similar to themselves. After the participants created their avatars, the RA started a session on \textit{Meta Horizon Workrooms} and placed the first two participants in a group workspace to complete the icebreaker exercise. At the same time, the third participant waited for them in a private virtual room. The RA added the third participant to the workspace after 10 minutes. In the virtual meeting room, each participant had a virtual monitor in front of them with their respective versions of the candidates' resumes. Participants could not see each other's monitors. They sat together at a virtual table, and the application randomly assigned them to a seat. Participants could use body movements and hands while talking to others. We cast the group workspace (Figure \ref{figure:vr-session}) and recorded audio of participants' conversations to analyze their decisions and identify their final candidate choice. 

We selected Meta Horizon Workrooms for this experiment for three reasons. First, it provided an accessible, high-quality virtual workspace that simulates real-world collaboration scenarios with multiple concurrent users and low latency. Second, it could connect computer monitors to each participant's headset using the Remote Desktop feature, allowing us to display the three different versions of the candidates' resumes. Third, this application enabled users to create their avatars quickly using Meta Avatars.   

\subsection{Procedure}
Participants attended our experimental sessions at our research laboratory. An RA was responsible for conducting the experiment and guiding the participants. At the beginning of the session, the RA explained the purpose of the study to the participants and answered their questions. The RAs emphasized that their participation was voluntary, that they were compensated, and that their responses were confidential. This stage lasted five minutes. The RAs provided a consent form to the participants and gave them time to read and ask questions. All participants had to consent to release their collected data to participate in this study; otherwise, the session would not have been conducted. We compensated participants recruited from the SONA pool with extra credit, while participants recruited through direct outreach and social media were compensated with \$20 electronic gift cards. Each session lasted 30 minutes on average.

The RA guided the participants to the check-in stations and assigned them nicknames so they could be recognized later for the surveys. The incumbents were nicknamed ``Cat'' and ``Dog,'' while the newcomer's nickname was ``Mouse.'' Each station had a label with the assigned nickname.

Participants completed a pre-treatment survey on Qualtrics, which included questions about their demographics (i.e., gender, ethnicity, race, and age), activity information (i.e., highest education level achieved, and current employment status), social self-efficacy (i.e., expectations to collaborate and form relationships while working together on a task), computational proficiency (i.e., how comfortable participants were using computers, phones, and VR). Since the study was conducted in the U.S., we also asked whether participants identified as Hispanic/Latino. Additionally, we ensured that participants did not know each other before the experiment. Participants sat at separate desks, divided by portable curtains, to prevent them from seeing each other's responses.

After completing the initial survey, the RA led the participants to the rooms according to the experimental conditions. Two participants completed the ice-breaker task while a third waited. After the ice-breaker task was completed, the third participant joined the other two participants for the hidden-profile task. We asked participants to introduce themselves again before starting the task. We gave the participants ten minutes to complete the task and select one candidate. 

The RA returned to the participants after the task was completed and relocated them to the check-in stations. Each participant completed a post-treatment survey on Qualtrics to assess their experiences working with the team. The participants had to evaluate their experience with the team and their teammates. To ensure that the participants recognized the correct team member in each question, we framed the questions about the incumbent as \textit{``For the partner you met first...''} and the ones regarding the newcomer as \textit{``For the partner you met later...''}. The third member had a different version of the survey and had to answer the questions based on the incumbents' nicknames. After participants had completed this final survey, RAs confirmed their completion, provided compensation, and conducted a brief debriefing of the experiment.

\subsection{Measurements}
We validated participants' responses to ensure they were reliable by calculating Cronbach's ($\alpha$) alpha. This test provided values above 0.75 for all scales, confirming their internal validity and allowing us to move on to the next steps of our data analysis. We provide the citations, reliability scores, and example items in Table \ref{tab:final_survey}.

\subsubsection{Dependent Variables}
\paragraph{Closeness}
We asked participants how closely they perceived each team member in their respective environment. Participants evaluated their closeness with each team member using a 7-point Likert scale using the items tested by \cite{Gachter2015}, which included questions from the `Inclusion of the Other in the Self' (IOS) Scale, the `Social Connectedness Scale,' and the `We Scale.' After checking that the items showed high-reliability levels ($\alpha=.90$), we averaged these items into a single score per participant. We defined this average as the closeness score ($C_{i \rightarrow j}$) to assess participant $i$'s closeness feelings to participant $j$. This score measured how close the participants felt to one another after completing the task.

\paragraph{Incumbents' Familiarity Bias}
Using participants' assessments of their relationships, we measured how different each incumbent felt connected with their incumbent compared to their newcomer. We operationalized the familiarity bias of each incumbent $i_1$ as the difference between their perceived closeness to the other incumbent $i_2$ (i.e., $C_{i_1 \rightarrow i_2}$) and their perceived closeness to the newcomer $n$ (i.e., $C_{i_1 \rightarrow n}$). This score ranged from -1 (i.e., feeling connected with the newcomer only) to 1 (i.e., feeling connected with the incumbent only). To normalize these closeness scores between participants and account for individual differences in rating scales, we calculated the relative difference from the absolute change between the closeness to $i_2$ and the closeness to $n$, and divided by the closeness to $i_2$ (Equation \ref{eq:familiarity-bias}). We calculated this familiarity score for each incumbent, giving us two scores per team (i.e., $i_1 \rightarrow i_2$ and $i_2 \rightarrow i_1$). 

\begin{equation}
FB_{i_1} = \frac{\left(C_{i_{1} \rightarrow i_{2}} - C_{i_{1} \rightarrow n}\right)}{C_{i_{1} \rightarrow i_{2}}}
\label{eq:familiarity-bias}
\end{equation}

\subsubsection{Independent Variables}

\paragraph{Experimental Conditions} We used a dummy variable to represent the experimental condition, where zero represented the In-Person sessions, and 1 represented the VR sessions. 

\paragraph{Perceived Similarity} We asked participants how similar they perceived themselves to their teammates. We used the items from \cite{brucks2022virtual}, including questions about their similarities with each member. 

\paragraph{Gender Homophily} We included a `same-gender' variable to verify that participants felt closer to each other because of gender homophily. 

\paragraph{Support provided by the Environment} To assess to what extent the environment supported participants to work on the task, we adapted questions from the \textit{Work Environment Satisfaction} scale \cite{tenorio2020syncmeet}. This scale includes questions measuring how participants feel about working in each environment, such as ``I could utilize all my skills and abilities to solve this task in this environment.''

\paragraph{Communication Metrics} Participants assessed how effective the communication with their partner was using the \textit{Communication Effectiveness Index} (CETI) \cite{lomas1989communicative}. This scale evaluates participants' ability to pay attention, communicate, respond, and interact. We also measured participants' communication using the \textit{Information Exchange} scale \cite{subramaniam2005influence}. This scale presents three statements asking whether the partner and the participant shared information, if they learned from each other, and if they exchanged ideas. 

\paragraph{Usability}
For the VR participants, we verified that using the headset and the workplace application was not an impediment to working on the collaborative task. Participants evaluated the system's usability using the SUS scale \cite{brooke1996sus}, consisting of 10 Likert-scale items with responses ranging from 1 (``Strongly disagree'') to 5 (``Strongly Agree''). SUS scores are calculated on a scale from 0 to 100, with higher scores indicating better usability.

\begin{table}[!htb]
\centering
\renewcommand{\arraystretch}{1.3}
\resizebox{\columnwidth}{!}{%
\begin{tabular}{p{0.1\textwidth}p{0.1\textwidth}p{0.25\textwidth}p{0.03\textwidth}p{0.03\textwidth}}
\toprule
\textbf{Dimension} &
  \textbf{Scale} &
  \textbf{Item/Question Example} &
  \textbf{Cit.} &
  \textbf{$\alpha$} \\ \midrule
Closeness &
  7-Point Likert Scale, 4 items &
  \textit{``Please select the appropriate number below to indicate to what extent you would use the term `WE' to characterize you and this partner''} &
  \cite{Gachter2015} &
  0.90 \\ \hline
Communication Effectiveness &
  5-Likert Scale, 10 items &
  \textit{``Please rate your partner's ability at... Indicating that he/she/they understands what is being said to him/her/them''} &
  \cite{lomas1989communicative} &
  0.89 \\ \hline
Information Exchange &
  7-Likert Scale, 3 items &
  \textit{``My partner and I learned from one another''} &
  \cite{subramaniam2005influence} &
  0.93 \\ \hline
Perceived Similarity &
  5-Likert Scale, 2 items &
  \textit{``How similar are you to your partner?''} &
  \cite{brucks2022virtual} &
  0.76 \\ \hline
Work Support by the Environment &
  5-Likert Scale, 2 items &
  \textit{``How efficiently did you communicate with your partners in this environment?''} &
  \cite{brucks2022virtual} &
  0.88 \\ \hline
Usability (VR participants) &
  5-Likert Scale, 10 items &
  \textit{``I thought the system was easy to use''} &
  \cite{brooke1996sus} &
  0.88
  \\ \bottomrule
\end{tabular}%
}
\caption{Final survey items}
\Description{Final survey items}
\label{tab:final_survey}
\end{table}

\subsubsection{Open-ended Questions}
Lastly, we included questions in the final survey to know more about the participants' impression of their teammates and working with them. We asked them, `Do you feel closer to one partner than another?' and `What were the most significant obstacles to relating with your partners?'

\subsection{Quantitative Analysis}
Using the participants' responses, we conducted the following statistical analyses using R 4.4.0 \cite{R2024}.  We first conducted Welch $t$-tests to check significant differences between participants' closeness scores based on the experimental conditions and relationships. We used this test since the samples in the experimental conditions had different sizes and variances. We also conducted a two-way ANOVA to test whether the experimental condition and the type of relationship (i.e., ``incumbent $\rightarrow$ incumbent,'' ``incumbent $\rightarrow$ newcomer,'' ``newcomer $\rightarrow$ incumbent'') had an effect on their perceived closeness. 

To test whether VR had a significant effect on incumbents' closeness to newcomers (H1) and newcomers' closeness to incumbents (H2), we conducted a mixed-effects linear regression analysis to estimate the factors that most explained their perceived closeness. Each observation was the evaluation of a participant on another. We modeled the sessions as random effects since each session had three participants. We used their reported closeness scores ($C_{i}$) as the variable to estimate and added the independent variables to our model. We included two interaction terms to verify whether the type of relationship moderated the effect of the experimental condition on perceived closeness. We used the package \texttt{lme4} \cite{Bates2015} to create these mixed-effects models.

We then conducted a Welch \textit{t}-test to check whether incumbents' familiarity bias differed significantly based on the conditions (H3). We also conducted a mixed-effects linear regression model to estimate the factors that most explained these biases and used sessions as random effects since each session had two incumbents. Given that the familiarity bias was operationalized between incumbents and newcomers, we also computed the relative change between the communication and perceived similarities among them.

Lastly, we conducted mixed-effects mediation models to examine whether participants' perceived similarity influenced the effect of using VR on perceived closeness (H4). We created different models to account for all the observations, as well as the directionality of the relationships (i.e., newcomers to incumbents, and incumbents to newcomers). We also modeled sessions as random effects since each participant evaluated two teammates. We created these models using the \texttt{mediation} package \cite{Tingley2014}, which allows moderation models with nested data structures.

For all the mixed-effects regression models, we verified the normality assumptions of these models and checked that multicollinearity was not an issue by calculating the models' covariates Variance Inflation Factors (VIF).

\subsection{Qualitative Analysis}
To gain deeper insights into participants' experiences, we conducted a qualitative analysis of open-ended survey responses, capturing their perspectives on challenges faced during the experiment and perceptions of newcomers and incumbents.

Using an iterative inductive approach \cite{saldana2021coding}, the first author conducted initial open coding of 10 responses (five incumbents and five newcomers), identifying 40 codes. Additional coding by the second and third authors expanded the code set, which was later consolidated into a 20-code codebook through affinity mapping \cite{nielsen2024affinity}. After coding 43 more responses collaboratively, the codebook was refined to 14 codes during team discussions. The finalized codebook was then shared among four researchers, who independently coded overlapping batches of responses (24 by the first researcher and 21 by each of the others). Collaborative discussions reconciled differences and highlighted key insights, with responses addressing multiple non-mutually exclusive codes.

The first and last authors conducted a thematic analysis \cite{braun2006using}, organizing codes into overarching themes and compiling relevant data for each. Drawing on the studies covered in Section \ref{literature_review}, they employed a deductive coding approach, iteratively refining themes over three collaborative meetings. This process clarified theme definitions and resulted in four main themes that provided insights into the experiences of both newcomers and incumbents.

\section{Results}
\label{results}
We recruited 87 participants, whose ages ranged from 18 to 43 years old, and included undergraduate students, graduate students, and staff members of our institution. Of these, 33 identified as male and 54 as female; 58 identified as White, 10 as Asian-American, 10 as African-American, and nine as ``Other.'' Fifteen participants identified as Latino or Hispanic. Most participants reported being more comfortable with Zoom (96\% feeling comfortable or very comfortable) than VR (32\% feeling comfortable or very comfortable). 

From this participant pool, we conducted a total of 29 sessions between April and September 2024: 14 IP sessions and 15 VR sessions. VR participants evaluated the system's usability with a mean SUS score of 69.42 ($SD=13.81$), indicating close to average usability based on the industry standard benchmark of 68 \cite{Lewis2018}. Lastly, we did not find a significant difference in the proportion of teams that answered correctly between the In-Person (IP) condition ($M=0.79, SD=0.43$) and the VR condition ($M=0.73, SD=0.46$). Table \ref{tab:descriptive} provides a descriptive summary of the data categorized by experimental conditions. 

\begin{table*}[!ht]
\centering
\small
\begin{tabular}{@{}l|ccc|ccc|cc|ccc@{}}
\toprule
& \multicolumn{3}{c}{\textbf{In Person}} & \multicolumn{3}{c}{\textbf{VR}} & \multicolumn{2}{c}{\textbf{\textit{t}-test}} & \multicolumn{3}{c}{\textbf{Total}} \\ \midrule
\textbf{Metric} & $N$ & Mean & SD & $N$ & Mean & SD & Statistic & \textit{p}-value & $N$ & Mean & SD \\ 
  \midrule
  Proportion of female members & 42 & 0.60 & 0.50 & 45 & 0.64 & 0.48 & -0.47 & 0.64 & 87 & 0.62 & 0.49 \\ 
  Age & 42 & 20.40 & 3.81 & 45 & 21.69 & 4.69 & -1.41 & 0.16 & 87 & 21.07 & 4.31 \\ 
  Proportion of Latinos/Hispanics & 42 & 0.14 & 0.35 & 45 & 0.20 & 0.40 & -0.70 & 0.48 & 87 & 0.17 & 0.38 \\ \midrule
  Closeness & 42 & 3.11 & 1.27 & 45 & 3.50 & 1.63 & -1.76 & 0.08 & 87 & 3.31 & 1.48 \\ 
  Perceived Similarity & 42 & 3.21 & 0.84 & 45 & 3.50 & 0.89 & -2.26 & 0.02 & 87 & 3.36 & 0.87 \\ 
  Communication Effectiveness & 42 & 3.78 & 0.70 & 45 & 3.80 & 0.84 & -0.13 & 0.90 & 87 & 3.79 & 0.77 \\ 
  Information Exchange & 42 & 5.96 & 1.07 & 45 & 6.03 & 1.22 & -0.44 & 0.66 & 87 & 6.00 & 1.15 \\ 
  Support by the Environment & 42 & 5.47 & 1.00 & 45 & 5.12 & 1.21 & 2.08 & 0.04 & 87 & 5.29 & 1.12 \\ \midrule
  Incumbents' Closeness to Incumbent & 28 & 4.05 & 1.18 & 30 & 3.85 & 1.55 & 0.57 & 0.57 & 58 & 3.95 & 1.37 \\
  Incumbents' Closeness to Newcomer & 28 & 2.88 & 1.22 & 30 & 3.03 & 1.59 & -0.40 & 0.69 & 58 & 2.96 & 1.41 \\
  Incumbents' Familiarity Bias & 28 & 0.29 & 0.20 & 30 & 0.21 & 0.24 & 1.29 & 0.20 & 58 & 0.25 & 0.22 \\
  Newcomers' Closeness to Incumbents & 14 & 2.39 & 0.76 & 15 & 3.62 & 1.70 & -3.57 & 0.00 & 29 & 3.03 & 1.46 \\ \midrule
  Teams with the Correct Answer (prop.) & 14 & 0.79 & 0.43 & 15 & 0.73 & 0.46 & 0.32 & 0.75 & 29 & 0.76 & 0.44 \\ 
\bottomrule
\end{tabular}
\caption{Descriptive Results of the Study. We present the metrics in four categories across in-person and VR conditions: participants' demographics, participants' final survey metrics, incumbent and newcomer metrics, and the proportion of teams with the correct answer. The number of participants/sessions ($N$), means, and standard deviations (SD) are reported for both conditions, along with the results of Welch's \textit{t}-tests comparing the two conditions. We observed significant differences for Perceived Similarity ($p < 0.05$), Support by the Environment ($p < 0.05$), and Newcomers' Closeness to Incumbents ($p < 0.001$).}
\Description{Descriptive Results of the Study. We present the metrics in four categories across in-person and VR conditions: participants' demographics, participants' final survey metrics, incumbent and newcomer metrics, and the proportion of teams with the correct answer. The number of participants/sessions ($N$), means, and standard deviations (SD) are reported for both conditions, along with the results of Welch's \textit{t}-tests comparing the two conditions. We observed significant differences for Perceived Similarity ($p < 0.05$), Support by the Environment ($p < 0.05$), and Newcomers' Closeness to Incumbents ($p < 0.001$).}
\label{tab:descriptive}
\end{table*}

\subsection{Quantitative Results}
As expected, incumbents' closeness to the other incumbents was the highest score across all the relationships. While the IP condition's incumbents reported a mean closeness to their first teammates of 4.05 ($SD=1.18$), the VR condition's incumbents reported a mean closeness of 3.85 ($SD=1.55$). Using VR did not make any statistically significant difference to how incumbents perceived other incumbents (Welch's $t=0.57, p>0.10$). 

In both conditions, incumbents reported lower closeness to their newcomers than to their incumbents. Participants in the IP condition reported a mean closeness of 2.88 ($SD=1.21$), while those in the VR condition reported a slightly higher mean of 3.03 ($SD=1.59$). However, this difference was not statistically significant ($t=-0.40, p>0.10$). 

Unlike the incumbents' reported closeness scores, we found that the VR condition's newcomers felt closer to their incumbents ($M=3.62, SD=1.70$) than the IP condition's newcomers ($M=2.39, SD=0.76$), a difference that was statistically significant ($t=-3.57, p < 0.001$). The effect size, measured using Cohen's $d$, was $d=0.91$, indicating a large effect. This result suggests that using VR significantly enhanced the newcomers' sense of closeness to incumbents compared to the IP condition.

\begin{figure}[!hbt]
    \centering
    \includegraphics[width=\columnwidth]{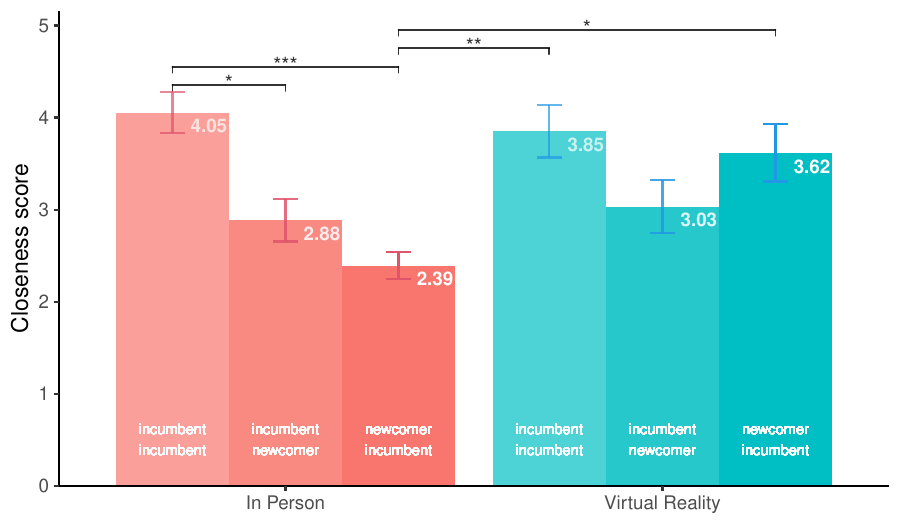}
    \caption{Participants' closeness per condition and relationship. Error bars represent standard deviations. Brackets represent statistically significant differences between two conditions using Tukey HSD tests ($p_{adj} < 0.05$). Each bracket shows its respective \textit{p}-adjusted value.}
    \label{fig:closeness-score-condition}
    \Description{Participants' closeness per condition and relationship. Error bars represent standard deviations. Brackets represent statistically significant differences between two conditions using Tukey HSD tests ($p_{adj} < 0.05$). Each bracket shows its respective \textit{p}-adjusted value.}
\end{figure}

Figure \ref{fig:closeness-score-condition} shows the reported closeness between incumbents, incumbents to newcomers, and newcomers to incumbents per condition. A two-way ANOVA test confirms the significant differences in closeness between types of relationships and conditions ($F(2,168)=9.28, p<0.001$). Although the experimental condition did not cause a statistically significant effect on the participants' closeness ($F(1,168)=3.46, p<0.10$), we found a significant interaction between the effects of the experimental condition and the type of relationship ($F(3,168)=4.2, p<0.05$). Post-hoc analysis confirmed that the incumbents felt strongly closer to the other incumbent rather than to the newcomer in the IP condition ($\Delta=1.17, p_{adj}<0.05$). Moreover, in the IP condition, the newcomers felt less close to the incumbents compared to how the incumbents felt to the other incumbents ($\Delta = 1.66, p_{adj} < 0.001$). Across conditions, we found that newcomers' closeness to incumbents was significantly higher in VR than IP ($\Delta=1.23, p_{adj}<0.05$). In sum, these results demonstrate the significant differences in participants' perceptions of closeness to others, showing that newcomers using VR felt closer to the rest of the group than those meeting in person. 

The mixed-effects linear regression model, accounting for participants' perceived similarity, gender homophily, communication patterns, and environmental factors, further confirms the influence of relationships and experimental conditions on closeness (Table \ref{tab:mixed-effect-closeness}). By using the relationship between incumbents as the baseline, we found that the closeness from incumbents to newcomers was significantly low ($\beta=-0.40,p<0.05$), as well as the closeness reported from newcomers to incumbents ($\beta=-0.63,p<0.001$). Regarding the other independent variables, we found that the participants' perceived similarity had a significant effect on their closeness ($\beta=0.38,p<0.001$). The first interaction term shows that the incumbents' closeness was not statistically higher in the VR condition ($\beta=0.19,p>0.05$). This indicates that VR did not lead to an increased sense of closeness for incumbents toward newcomers, providing no evidence to support H1.

The second interaction term shows that the newcomers' closeness was significantly higher in the VR condition ($\beta=0.49,p<0.05$), indicating that VR facilitated stronger connections between newcomers and their incumbents compared to the IP condition. This result supports H2, demonstrating that VR enhances newcomers' sense of closeness to incumbents.

We found that incumbents' familiarity bias was smaller in the VR condition ($M=0.21, SD=0.24$) than in the IP condition ($M=0.29, SD=0.20$). However, this difference was not statistically significant ($t=1.29, p>0.10$), suggesting that the medium in which participants met did not influence incumbents' attachment to the first member. Consequently, H3 is not supported.

\begin{table*}[ht]
\centering
\small
\begin{tabular}{lccccc}
  \hline
 & \textbf{Estimate} & \textbf{Std. Error} & \textbf{d.f.} & \textbf{\textit{t} value} & \textbf{Pr($>$$|$t$|$)} \\ 
  \hline
  \textbf{Fixed Effects} & & & & & \\
  Intercept & 0.26 & 0.18 & 58.68 & 1.43 & 0.16 \\ 
  Condition (1: VR) & -0.08 & 0.25 & 52.96 & -0.31 & 0.76 \\ 
  Relationship: incumbent $\rightarrow$ newcomer & -0.40* & 0.18 & 137.33 & -2.22 & 0.03 \\ 
  Relationship: newcomer $\rightarrow$ incumbent & -0.63*** & 0.18 & 139.10 & -3.51 & 0.00 \\ 
  Gender Homophily & 0.00 & 0.12 & 158.39 & 0.00 & 1.00 \\ 
  Perceived Similarity & 0.38*** & 0.07 & 146.16 & 5.13 & 0.00 \\ 
  Support by the Environment & 0.07 & 0.07 & 161.67 & 0.94 & 0.35 \\ 
  Comm. Effectiveness & 0.07 & 0.08 & 150.50 & 0.91 & 0.36 \\ 
  Information Exchange & 0.02 & 0.07 & 152.04 & 0.22 & 0.83 \\ 
  (H1) Condition $\times$ Relationship: incumbent $\rightarrow$ newcomer & 0.19 & 0.23 & 135.08 & 0.81 & 0.42 \\ 
  (H2) Condition $\times$ Relationship: newcomer $\rightarrow$ incumbent & 0.49* & 0.24 & 137.05 & 2.04 & 0.04 \\ \midrule
  \textbf{Random Effects} & & & & & \\
  Intercept (Variance) & 0.24 & 0.48 & & & \\
  Residual (Variance) & 0.38 & 0.62 & & & \\
   \hline
\end{tabular}
\caption{Mixed-effects linear regression model estimating participants' closeness with other teammates. We used the ``incumbent $\rightarrow$ incumbent'' as the baseline for relationships. Number of Observations: 174 (i.e., 87 participants evaluating two members). Groups (i.e., Sessions): 29. We computed the Pseudo-R-squared for Generalized Mixed-Effect models: Marginal $R^2=0.35$, Conditional $R^2=0.60$. Significance codes: * $p < .05$, ** $p < .01$, *** $p < .001$.}
\Description{Mixed-effects linear regression model estimating participants' closeness with other teammates. We used the ``incumbent $\rightarrow$ incumbent'' as the baseline for relationships. Number of Observations: 174 (i.e., 87 participants evaluating two members). Groups (i.e., Sessions): 29. We computed the Pseudo-R-squared for Generalized Mixed-Effect models: Marginal $R^2=0.35$, Conditional $R^2=0.60$. Significance codes: * $p < .05$, ** $p < .01$, *** $p < .001$.}
\label{tab:mixed-effect-closeness}
\end{table*}

\begin{figure}[!hbt]
    \centering
    \includegraphics[width=\columnwidth]{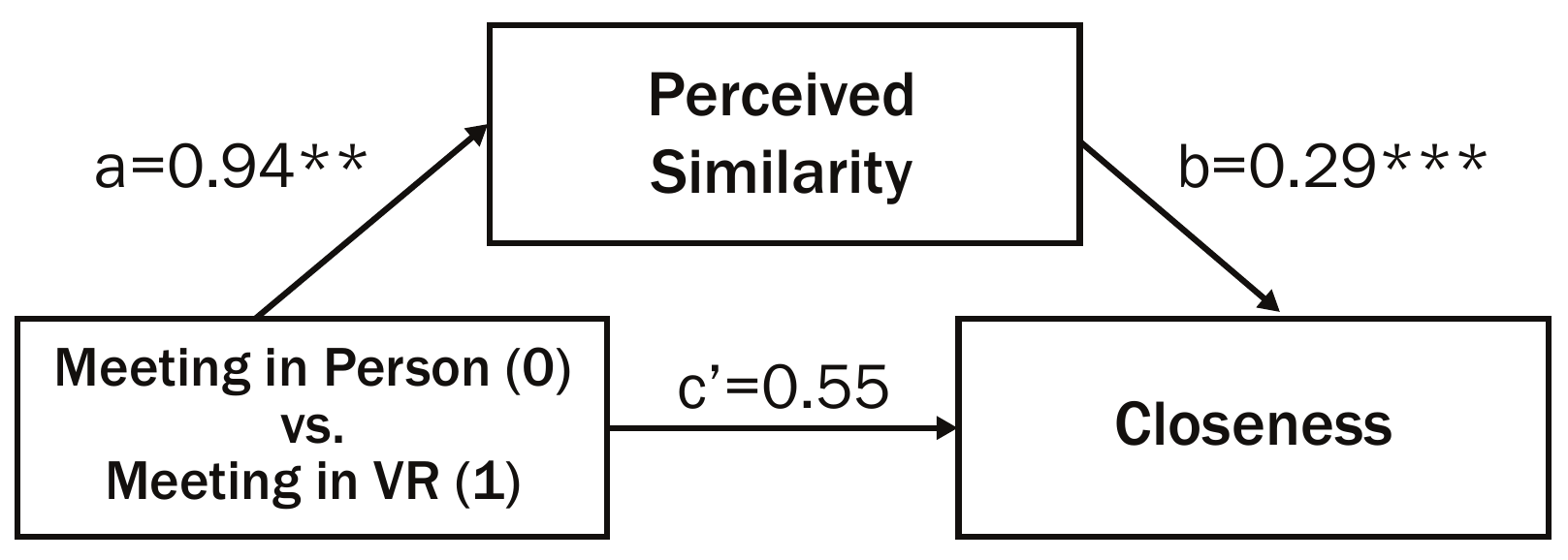}
    \caption{Mediation analysis showing that the relationship between the medium condition and the newcomers' closeness to incumbents is partially mediated by their perceived similarity. While the direct effect is $c'$, the indirect effect is $a \times b$.}
    \label{fig:mediation-analysis}
    \Description{Mediation analysis showing that the relationship between the medium condition and the newcomers' closeness to incumbents is partially mediated by their perceived similarity. While the direct effect is $c'$, the indirect effect is $a \times b$.}
\end{figure}

Lastly, we analyzed whether the effect of using VR on participants' closeness was mediated by their perceived similarity. Given the significant differences across the types of relationships, we first conducted a mixed-effects mediation analysis with the newcomers' closeness to incumbents. We found that their perceived similarities to the incumbents mediated 32\% of the total relationship between using VR and their closeness to the incumbents (Figure \ref{fig:mediation-analysis}). Consistent with the mixed-effects regression model, this mediation analysis indicates that the experimental condition did not have a significant direct effect on newcomers' closeness to incumbents ($c=0.55,p>0.05,$ Marginal Pseudo-$R^2=0.28$). However, we found that meeting in VR helped newcomers perceive themselves more similar to the incumbents than meeting in person ($a=0.94,p<0.01,$ Marginal Pseudo-$R^2=0.21$), and newcomers' perceived similarity had a significant effect on their closeness to incumbents ($b=0.29,p<0.001,$ Marginal Pseudo-$R^2=0.28$). This finding suggests that using VR enabled newcomers to feel more similar to the current team members, compared to meeting in person, which led to higher closeness feelings when working together. In the case of the incumbents' closeness to newcomers, we found no statistically significant relationships. VR incumbents did not feel more similar to their newcomers than IP incumbents, nor did their perceived similarities explain their closeness with newcomers. Therefore, H4 is partially supported. 

\subsection{Qualitative Results}

\subsubsection{Virtual Reality Shielding}
This theme addresses why newcomers in VR felt significantly closer to incumbents than newcomers in IP (supporting H2). Seventy-three percent (73\%) of the newcomers in VR reported a sense of psychological safety, enabling them to engage without fear of judgment or social repercussions \cite{edmondson2014psychological}. Meanwhile, 42\% of the participants in the IP condition reported amplified feelings of exclusion due to visible pre-existing connections among incumbents. For example, P33 mentioned how the incumbents were already focused on the task without considering them: \textit{``I didn't want to interrupt. They started talking about the task, and it was a challenge to contribute when I didn't even know their names.''}. Similarly, P24 described feelings of shame and the physical manifestation: \textit{``I felt a little bit of shame, and I turned red, the situation was awkward.''} 

By contrast, 11 of the 15 newcomers in the VR condition reported that the VR setting helped them feel more comfortable and closer to the incumbents. Participants described how the VR application acted as a `shield,' enabling them to focus on the task rather than potential social anxieties. For example, P21 described how the avatar's cartoonish appearance contributed to a more relaxed social environment: \textit{``I didn't face any challenge, the avatar looked funny, and my team members started directly to work on the task once I entered the session.''} Another newcomer (P6) explained that they had feelings of nervousness about joining the session but how the customization of the avatar relieved this feeling: \textit{``At the beginning, I was nervous because I was going to enter the session, but I enjoyed customizing my own avatar, which made me feel more excited about the activity.''} This sentiment suggests that the VR application allowed participants to engage with the activity in a neutral and focused manner.

Moreover, the mediated nature of communication in VR, combined with the abstraction provided by avatars, allowed newcomers to participate without the social pressures often present in face-to-face interactions. For instance, newcomer P15 described how the environment helped them feel less self-conscious when speaking: \textit{``I felt like I could just speak up without overthinking. I didn't feel nervous at all.''}. These comments illustrate how the VR application provided a layer of psychological safety, enabling newcomers to engage more freely with incumbents. This might have facilitated a more inclusive atmosphere, where the emphasis was on task completion rather than on social hierarchies or personal judgments.

\subsubsection{The Impact of Time Over Medium}
Nevertheless, incumbents in the VR condition did not experience the same ease in connecting with newcomers (rejecting H1) as they did with fellow incumbents (rejecting H3). Among the 30 incumbents, 30\% expressed discomfort with the abrupt transition to the task after the newcomer's arrival. For instance, P8 described feeling uncomfortable when the newcomer joined, and the group had to immediately shift focus: \textit{``The [newcomer] joined the session, and we had to jump straight into the activity. It felt awkward because we were talking about other things when she joined.''} Moreover, 56\% of incumbents expressed that features of the VR environment, such as the avatars' appearance and the application's immersiveness, did not significantly impact their sense of closeness to newcomers, as their connection with other incumbents remained stronger. As participant P33 explained: \textit{``It was harder to collaborate with the [newcomer] because he was so quiet, the [incumbent] was much more active and we already talked about other things.}''

In both experimental conditions, 13 incumbent participants emphasized that time was the primary factor in developing a sense of closeness with other incumbents. Having more time to interact before the newcomer joined allowed incumbents to establish a prior relationship, reinforcing their connection. In contrast, the limited time available to engage with the newcomer hindered the development of a similar bond, making it more challenging to establish the same level of closeness. For example, P4 mentioned: \textit{``I had more time to interact with the [incumbent] and established a connection.}'' This finding suggests that closeness among participants was primarily built through longer interactions rather than initial visual cues. Neither physical appearance in the IP condition nor avatar appearance in VR significantly influenced their sense of closeness. Instead, the duration of interaction played an important role in developing familiarity and strengthening a sense of closeness. Notably, VR did not offer a significant advantage in mitigating familiarity bias, as the medium itself did not compensate for the incumbents' pre-existing relationships formed through prior time spent together.

\subsubsection{Overcoming Non-Verbal Constraints Through Kinesthetic Cues}
The participants' responses also underscore how using VR affected their perceptions of each other, ultimately affecting how similar or different they felt with each other (partially supporting H4). Compared to participants working IP, participants in VR reported that most of their attention went to the coexistence of physical and virtual elements. While physically situated in their own rooms, participants were tele-transported to a shared virtual workspace, creating an immersive and engaging experience. Structured elements, such as avatar placements and the design of the virtual environment, played a crucial role in reinforcing this presence, as specifically noted by four participants. For instance, P22 described: \textit{``I felt completely immersed because I was in an office, and all of us were placed in seats.''} Beyond the virtual environment itself, seven participants (15\%) also highlighted the role of real-world objects in enhancing their sense of presence. They appreciated the blend of physical and virtual elements, noting how real-world objects like desks and computers complemented the virtual setting. This interplay reinforced the feeling of being in a shared workspace, as P16 explained:\textit{``I was in the virtual office, but I felt comfortable because it felt like I was really there with the team.''} 

Moreover, one-third of the participants emphasized the immersive nature of VR. Among them, six newcomers mentioned how seeing their avatars replicate their own movements made them feel more comfortable and helped them quickly adapt to collaborating with the team. One newcomer (P3) shared: \textit{``I noticed that I was moving my hands, and my avatar was doing the same. That made me feel immersed, and it's the reason I did not feel weird during the activity.''} Another participant (P12) mentioned that the rotation of the avatars when someone was referring to themselves was helpful to stay engaged in the conversation: \textit{``The avatars of the other team members were facing me, and I also noticed that my avatar's hands were mirroring my movements. Because of that, I did not even notice the lack of facial expressions.''} In contrast, participants who worked IP paid less attention to their environment and were more conscious of the incumbents who were working on the task. For instance, a newcomer working in the IP condition (P7) said \textit{``they [incumbents] were able to connect beforehand, and I was not able to talk or get to know them. They felt like they knew each other very well, which made me uncomfortable sharing my ideas to a full extent.''}

Another main explanation of this difference between VR and IP was the absence of facial expressions and other traditional non-verbal cues in VR, which was reported by 71\% of the VR participants. These features and cues are crucial for interpreting emotions, intentions, and perceived differences. This limitation likely influenced how incumbents and newcomers perceived each other's similarities and how closeness developed within the team. Two newcomer participants stated that the simplified and cartoonish avatars helped them focus on both the task and their conversations with others. For example, P18 noted that seeing another participant's avatar smile created a sense of excitement, making them feel more comfortable during the activity: \textit{``It was curious how some [incumbent] smiled, and I felt excited.''}

Despite the benefits of this design, 32 of the 45 participants in the VR condition described how the lack of facial expressions made communication more demanding and, at times, distracting. For example, P19 shared that trying to interpret their teammates' emotions through avatars required extra effort, which took their focus away from the task: \textit{``I couldn't see their faces, so I didn't focus on the conversation as much as I could have. I was trying from the beginning to get some insights from the avatar faces.''}. Other participants expressed frustration in not having facial expressions since they were very important to understanding what others were thinking: \textit{[P10] ``Not seeing their facial expressions was hard for me as I wasn't able to gauge what they were thinking as well as if it would've been in person.''}

To overcome these challenges, 19 participants in the VR condition adapted by using alternative kinesthetic cues, such as gestures or avatar interactions, which played a significant role in fostering a sense of connection. For instance, P42 described how a playful interaction helped them feel included: \textit{``We discovered that we could do a high five between team members, and that was so fun. I didn't feel like I was a stranger. I think I didn't have any challenges collaborating with them.''} Similarly, P29 noted how simple gestures, like a thumbs-up, helped build comfort and connection: \textit{``Some of the team members gave me a thumbs up, and that was nice. From that moment, I felt comfortable with the team.''}

\subsubsection{Blurring Realities: The Impact of Physical and Virtual Cues on Presence}
Our last theme provides additional context for understanding the nuanced effects of VR on team dynamics, highlighting how the environment influences behavior. More than half of the participants (53\%) found the virtual office highly immersive, describing it as feeling \textit{``almost real.''} However, thirteen participants (28\%) described the odd sensation of being physically alone yet visually and interactively present with others in the virtual space. The presence of real-world objects, such as their computer or desk, served as grounding elements that reinforced their physical reality but also intensified the artificial nature of the virtual one. As one participant (P52) shared: \textit{``I was in the virtual office, but I felt weird because, at the same time, I was alone with my computer in an isolated office. That made me feel kind of strange.''} Another incumbent (P26) elaborated on this feeling of detachment caused by the overlap of real and virtual elements: \textit{``It was a challenge because I was able to see the text of the activity on my real computer while also being in a virtual office. That made me feel strange, like a sense of non-presence but presence at the same time.''}

This theme characterizes the duality experienced by participants as they navigated the coexistence of physical and virtual environments in VR. While participants were physically located in their individual offices, the immersive nature of VR transported them to a shared virtual workspace. This juxtaposition of the tangible and the simulated created a blend of presence that was both engaging and disorienting.

\section{Discussion}
\label{discussion}
In this study, we examined how VR could promote team members' closeness when new members join an existing team. We conducted a between-subject experiment with teams meeting in one of two different environments---Virtual Reality (VR) and In-Person (IP)---and studied how incumbents and newcomers perceived each other by measuring their perceived closeness and the resultant familiarity bias. We found a positive and statistically significant effect of using a VR application on newcomers' closeness to incumbents. However, we found no significant evidence supporting that the VR setting positively affected incumbents' closeness to newcomers, and there was no significant difference in mitigating their familiarity bias. The participants' open-ended responses highlight how VR provided a shield to newcomers, fostering their safety and interactions with their incumbents. This section discusses how our findings advance understanding of VR in collaborative settings.

\subsection{Asymmetric Effects of Virtual Reality}
Our RQ1 asked whether using VR when forming teams influences the perceived closeness of incumbents to newcomers, and RQ2 focused on newcomers' closeness to incumbents. We found that newcomers in the VR condition perceived themselves more similar to their incumbents than the newcomers working IP, which mediated the positive effect of the VR condition on newcomers' closeness. While our findings reveal that the VR application fostered newcomers' perceptions of closeness to incumbents, incumbents' closeness to newcomers remained unchanged across both experimental conditions. Regardless of the setting, incumbents reported feeling neither similar nor particularly close to newcomers. This result extends prior research in HCI, showing that online communication systems can affect team members' perceptions in different ways \cite{yee2007proteus}. The current study identifies the asymmetric effects of employing VR on participants' perceived closeness. While VR has been shown to facilitate co-presence and improve the overall interaction experience in collaborative settings, our findings indicate that these benefits do not necessarily extend to all users equally.

Moreover, we asked whether VR could mitigate incumbents' familiarity bias against newcomers (RQ3). Although it was reduced in the VR application with respect to the IP setting, this difference was not statistically significant. This result could be attributed to incumbents' pre-established team dynamics. When the session began, incumbents had already had time to engage with their teammates and orient themselves to the task. By the time the newcomer arrived, incumbents were already in a task-focused mindset, leaving little room for social interactions that would foster a connection with the newcomer. This observation aligns with prior findings that highlight the importance of early social interactions in building team rapport \cite{10.1145/2998181.2998300, sawyer2010social}. Without deliberate interventions, VR alone does not seem sufficient to disrupt incumbents' reliance on pre-existing relationships or to encourage active engagement with newcomers.

\subsection{Virtual Reality: A Safe Space for Newcomers}
These findings highlight how VR can be employed to promote a more inclusive platform for newcomers. This study extends existing research on team familiarity \cite{10.1145/2998181.2998300, muskat2022team}, emphasizing the challenges newcomers face in navigating pre-existing relationships and biases within teams. Previous research indicates that newcomers often experience social pressure and anxiety when integrating into an established team \cite{kraut2010dealing, choi2004minority}. Our study suggests that VR can alleviate some of these challenges by making team members look more similar to each other. Features such as avatars and immersive spaces can help reduce the intensity of social cues that amplify social anxieties. This ``virtual shielding'' effect enabled newcomers to focus on collaboration and task completion rather than navigating the interpersonal complexities of team integration. The VR application offered a comfortable platform for newcomers to engage with established team members, making it a valuable tool for fostering equitable and effective team interactions.

Extensive research highlights the importance of technology's impact in making individuals' differences visible \cite{gomez2020impact,carte2004capabilities}. Our study extends these previous studies by showing how VR can reduce the visibility of demographic or cultural differences through customizable, similar avatars and neutral settings. VR applications in team settings should consider how the avatar design can promote team members' inclusion, supporting diverse team integration and shifting the focus toward shared goals and contributions. Yet, the lack of evidence supporting the idea that VR positively impacts existing team members underscores the need for alternative strategies to enhance this integration process. Approaches such as redesigned onboarding procedures or pre-training programs may have a more significant effect on fostering closer relationships between incumbents and newcomers. Previous research in HCI has demonstrated that socio-technical systems can facilitate team formation through short, structured interactions \cite{umbelino2021prototeams, lykourentzou2017team}. By orchestrating social introductions and interactions, VR could offer new strategies to address this challenge. For example, ``speed dating'' interactions could rotate team members through brief conversations, allowing them to learn about each other quickly. Additionally, team-based games requiring physical collaboration could create shared experiences that help mitigate familiarity bias. Even within the VR setting, exploring innovative methods to improve team integration and strengthen interpersonal connections between incumbents and newcomers will continue to be essential for VR adoption.

\subsection{Design Implications}

\paragraph{Altering Physical and Social Norms in VR}
While it is valuable to bring effective aspects of real-world collaboration into VR, designers should leverage the unique capabilities of VR to overcome the limitations of physical environments and strengthen relationships among team members. For example, VR allows for customizable avatars that can minimize physical appearance-based biases by focusing on shared roles or expertise rather than individual traits \cite{GatherTown, FrameVR}. Similarly, controlled non-verbal cues, such as highlighting active speakers or enabling customizable gestures, can ensure clearer communication and equitable participation, especially for individuals who might feel overshadowed in traditional settings. VR can also create blended interaction spaces, such as collaborative environments that adapt dynamically to task needs, enabling teams to move seamlessly between brainstorming, task execution, and social bonding without the constraints of a static physical location. Therefore, VR can be a potential tool that offers the opportunity to create such spaces, breaking down traditional barriers and facilitating the seamless integration of newcomers into teams. 
 
\paragraph{Integrating Kinesthetic and Interactive Features}
Our findings highlight the importance of considering richer approaches to avatar representation beyond improving facial expressions. While expressiveness is valuable and much research has focused on enhancing the realism of avatars \cite{waltemate2018impact, oh2016let}, other aspects of avatar design and interactions can be tailored specifically to address team formation needs, such as fostering inclusion and reducing familiarity bias. One design direction is enhancing the kinesthetic and interactive features of avatars. For example, incorporating haptic feedback, such as a vibration when team members ``high five'' or shake hands, can simulate physical contact and foster a sense of camaraderie. Additionally, enabling shared object manipulation, like collaboratively moving a virtual whiteboard or assembling a puzzle, can promote teamwork and a shared sense of purpose. Body tracking for natural gestures such as nodding, waving, or leaning in and spatial audio integration can further enhance the sense of presence and connection between team members \cite{abbas2023virtual, li2021social, williamson2022digital}.

\paragraph{Team Formation and Team-Oriented Avatar Customization}
Lastly, collaborative VR applications could serve as a valuable tool for supporting team formation. Masking individuals' physical appearances could enable teams to focus on building relationships and getting to know one another \cite{Whiting2020}. Team members can later transition to meeting in person or using other modalities to continue their collaboration. Moreover, avatar customization in VR applications offers an opportunity to align user representation with the specific dynamics of team collaboration. Rather than emphasizing purely individual personalization, these features could be designed to reflect team-oriented aspects, such as shared goals, complementary skills, or collaborative roles. For instance, teams can better identify each member's role and contributions by enabling users to tailor their avatars with visual markers such as professional backgrounds, representations of expertise, or elements reflecting shared interests.

\subsection{Limitations and Future Work}
While our study provides valuable insights, it is important to acknowledge its main limitations. One key limitation is the demographic homogeneity of our participants, who were primarily from a similar age group and educational background. This limitation raises questions about the generalizability of our findings to more diverse populations. Future research should aim to include a more varied participant pool to explore how different cultural and demographic factors influence team familiarity in virtual environments. Second, a larger sample size could increase the statistical validity of our study. Future research should replicate and increase the sample size to obtain results with higher significance levels. 

Third, several experimental design choices could have introduced bias to our results. For example, we did not provide an ice-breaker question to the third participant, which could have also explained the high attachment to the first partner. Despite this choice in the design of the experiment, the changes in closeness from incumbent to newcomer would have remained similar between the two experimental conditions. Another example is the short time that incumbents had to meet during this brief experiment and the short task that the team was required to do. While experiments conducted in laboratories can offer internal validity, these short-term teams have no past and no future. Future studies should examine real teams that are deciding to include new members. Additionally, longitudinal studies could provide deeper insights into how team members' closeness evolves in teams that regularly collaborate in VR settings.

Another limitation is the specific nature of the task used in our experiment, which involved decision-making based on shared information. Future studies could explore a broader range of tasks, such as creative tasks, to see how team members' closeness is impacted by a different type of task that could involve more discussion of ideas between members. Using only one VR application was also a limitation, as other applications may have offered different features and affected individuals differently. Future work should consider other VR applications as well. Finally, as VR technology continues to evolve, it will be necessary to re-evaluate how different the results could be when using more advanced and usable VR devices. Future research could explore how advancements in VR, such as improved sensory feedback and haptic technology, might further enhance or alter the dynamics of team interactions in virtual environments.

\section{Conclusion}
\label{conclusion}
This study explored how meeting in Virtual Reality can affect closeness and familiarity among team members. We examined empirical data about the perception of closeness between incumbents and newcomers through a controlled, between-subjects experiment conducted in both In-Person (IP) and Virtual Reality (VR). Our findings demonstrate how employing VR affects team members' closeness in asymmetric ways. While newcomers in VR felt closer to their teams than newcomers working in person, incumbents did not feel significantly different toward their newcomers across conditions. The findings suggest that VR could abstract participants' appearances, resulting in higher closeness toward existing team members. 

As teams and organizations increasingly adopt hybrid and remote collaboration tools, HCI researchers and practitioners will play an important role in bridging social gaps and promoting equal participation by leveraging online communication systems. Future research and development should focus on enhancing these virtual environments to improve teams' cohesion and performance further. We hope this work will inspire future studies and systems to continue exploring innovative ways to integrate VR into collaborative practices, ultimately creating more cohesive and effective teams in virtual settings.

\begin{acks}
This work was supported by the Alfred P. Sloan Foundation (G-2024-22427) and the University of Notre Dame's Lucy Family Institute for Data \& Society. We also thank the anonymous reviewers for their feedback and suggestions.
\end{acks}

\bibliographystyle{ACM-Reference-Format}
\bibliography{main}

\appendix

\section{Ice-breaker questions}
\label{appendix:ice-breaker}
Please go through these questions together, you do not need to answer them all.
\begin{itemize}
    \item Share [University Name] Introductions 
    \item What's the last TV show or movie you watched and enjoyed?
    \item Do you have any pets, and if not, what kind of pet would you like to have?
    \item If you could travel anywhere in the world, where would you go and why?
    \item What was your dream job as a kid?
    \item What's your favorite type of food, and why do you love it?
    \item What's a hobby or activity you enjoy doing in your free time?
    \item What kind of music do you like to listen to, and do you have any favorite artists?
    \item What's a skill or talent you wish you had, and why?
    \item What's the best piece of advice you've ever received, and did you follow it?
\end{itemize}

\end{document}